\documentclass[twocolumn,prl]{revtex4}
\RequirePackage{xspace}
\usepackage{graphicx}
\def\be{\begin{equation}}
\def\ee{\end{equation}}
\def\bearr{\begin{eqnarray}}
\def\eearr{\end{eqnarray}}
\def\tc{T$_c~$}

\def\c2{CuO$_2~$}

\def\lsco{LSCO~}
\def\lco{La$_2$CuO$_4$~}
\def\lbco{La$_{2-x}Ba_x$CuO$_4$~}

\def\ndc{Nd$_{2-x}$Ce$_x$CuO$_4$~}
\def\nd{Nd$_{2}$CuO$_4$~}
\def\prc{Pr$_{2-x}$Ce$_x$CuO$_4$~}

\def\lar{La$_{2-x}^{3+}$R$^{3+}_x$CuO$_4$~}
\def\lbco{${\rm La_{2-x}Ba_xCuO_2}$~}
\def\lczno{${\rm La_{2}Cu_{1-x}Zn_xO_2}$~}

\def\ce3{Ce$^{3+}$~}
\def\cut{Cu$^{2+}$~}

\def\cu1{Cu$^{1+}$~}
\def\cu2{Cu$^{2+}$~}

\def\dx2{Cu 3$d_{x^2-y^2}$~}
\begin{document}
\preprint{IMSc-2005/05/14}

\title{A resolution to doping asymmetry puzzle in high \tc cuprates}

\author{ G. Baskaran \\
Institute of Mathematical Sciences, C.I.T. Campus, Chennai 600 113, India}
\begin{abstract}
We present a microscopic model for `electron doped'  \ndc family and 
offer a resolution to a long standing doping 
asymmetry puzzle. Here, i) Ce atoms do not dope free electrons, instead 
a Ce atom effectively quenches a \cut spin moment at an adjacent site, 
at an energy 
scale $>$ superexchange J of CuO$_2$
  plane and `site dilutes' the Mott insulator.
ii) effective chemical pressure, caused by increased Ce substitution, induces a
{\em first order Mott insulator to superconductor transition} in the 
CuO$_2$ plane. We predict,
i) {\em equal number of +e and -e carriers} in metallic state and 
ii) a line of first order 
transition ending at a critical point in normal state. Phenomenology 
gets organized.

\end{abstract}
\maketitle
Following the seminal discovery of superconductivity in \lbco\cite{bednorz}, 
the idea of resonating valence bond mechanism of superconductivity in `hole doped
Mott insulator' was born\cite{pwascience}. In this theory\cite{bza+},
based on a single band model, electron doping
of Mott insulator should equally well lead to superconductivity and one
expects some kind of symmetry between electron and hole doping. 
When a so called `electron doped' superconductor \ndc was discovered\cite{Takagi}
there was a surprise. At the level of phase diagram (figure 1a) there was a striking 
asymmetry and significant differences. A large body of experiments[4-23] devoted 
to a study of this asymmetry  exists now. In spite of several theoretical 
\cite{theory} attempts, our understanding of this asymmetry
is not satisfactory.

Antiferromagnetism in \ndc family extends to a doping of nearly 12 \%, 
in contrast to a 2\% doping in \lsco (figure 1).  Superconductivity 
appears abruptly at a first order transition with a maximum \tc which 
decreases on further Ce doping\cite{Takagi,Muon,keimer}. 
Hall and other transport measurements show 
anomalies\cite{transport,greene1}, different from that shown by hole 
doped cases. 

The large asymmetry naturally points to some missing features
in the current modeling of superconductivity in \ndc family.
In the present article we identify these features and propose a 
new model, where strong electron correlations continue to play
major role. The missing features follow from some known experimental 
results and some physical arguments. Our model is consistent with the 
large body of existing phenomenology. It also makes some predictions. 

We argue that root cause of doping asymmetry lies in the
T'-structure of \ndc family and ambivalent (sic) character of Ce valence.
T' differs from T, the structure of \lco: apical oxygen of CuO$_4$ 
octahedra (T-structure) are displaced 
in the ab-plane and fall in line with the oxygens of CuO$_2$ layers along
the c-axis. A remarkable consequence of missing apical oxygen in 
T'-structure seems 
to be a nearly 1 eV reduction in Mott-Hubbard gap in \ndc (compared
to \lco), as seen in recent x-ray scattering experiment\cite{MHgap}.
This makes \nd a fragile Mott insulator, close to a Mott insulator to 
superconductor transition point. The unstable tetra valent state
of Ce is likely to play a key role in quenching a Cu$^{2+}$ spin moment,
and also generate required chemical pressure for a Mott insulator to
superconductor transition, that we propose below.
\begin{figure}
\includegraphics[width=8.5cm]{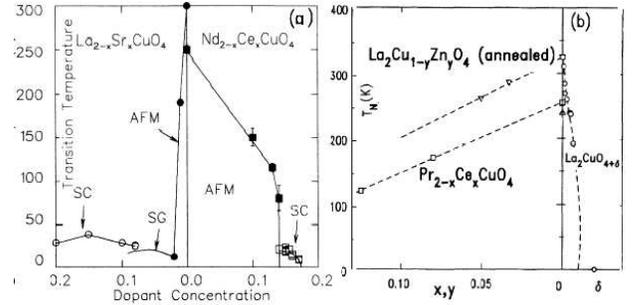}
\caption{\label{Fig1}a) Experimental phase diagram of electron and hole doped
cuprates, reproduced from ref\cite{Muon}. b) doping dependence of Neel temperature
for electron doped \prc, 
oxygen doped ${\rm La_2CuO_{4+\delta}}$ and Zn doped ${\rm La_{2}Cu_{1-y}Zn_yO_2}$, 
reproduced from reference\cite{keimer}.}
\end{figure}

In what follows, we define our microscopic model and discuss its 
solution. We also indicate how existing quantitative theoretical 
understanding on the hole doped case can be transfered to the \ndc
family, because of some special feature of our model.

Hypothesis leading to our model are: A) 
CuO$_2$  planes are not electron doped by Ce reservoir, instead every
Ce effectively quenches a \cut spin moment of an adjacent site, over an energy 
scale $>$ superexchange J of CuO$_2$ plane and `site dilutes' the Mott insulator.
B) chemical pressure  arising from increased Ce substitution, induces a
{\em first order Mott insulator to superconductor transition} in the 
CuO$_2$ plane, resulting in a `{\em self doping}' of Mott insulator with 
{\em equal number of +e and -e carriers}. 

Our model is `a site diluted Heisenberg model' for the insulating
phase, which exists in the experimental up to a value of 
$x_c \approx 0.12$. For $x > x_c$, we have a conductor, described 
by a `site diluted 2-species t-J model', which we elaborat later. 
In both cases, number of `missing Cu sites' or `diluted sites' is $Nx$,
where N is the number of sites and $x$ is the Ce density. 

It should be mentioned that effective `site dilution' as an 
experimental consequence of Ce doping (in the insulating 
antiferromagnetic order regime) is known in the literature\cite{keimer}, 
including a recent theoretical fit of their neutron scattering result 
to the site diluted Heisenberg model by Mang et al.\cite{Lynn}. 
Results, which is behind the site dilution hypothesis are:
i) the Neel temperature T$_{\rm N}$ has a slope which is the
same as the slope for a well established site dilution situation
namely \lczno, and ii) the long range antiferromagnetic order 
continues to be a commensurate $(\pi,\pi)$ order.
These two results are qualitatively different from the hole doped case,
where T$_{\rm N}$ is suppressed completely by about 2 \% doping 
and any magnetic quasi long range order is incommensurate and 
not at $(\pi,\pi)$.
We take site dilution as a phenomenological input to our modeling
and do not discuss mechanism and theory behind it. As far the theory
of site diluted 2D Heisenberg model, we refer to existing work in
the literature.

Two sceneries are possible in metallic state ($x > x_c$) of \ndc: 
i) Ce does not quench a Cu$^{2+}$ moment (i.e., no site dilution), 
instead becomes Ce$^{4+}$ and electron dopes the CuO$_2$ layer, and 
ii)  spin quenching (site dilution) continues and we reach 
metallic state through a Mott transition, induced by chemical 
pressure. Recent observation of Kondo  like effect\cite{kondo} in 
\prc, according to us, indicates that Ce continues to be involved 
in spin quenching in the 
metallic state. Further there is a strong indication of first order 
insulator to metal transition\cite{phaseSep}, reminiscent of of 
ET-salts (organic 
superconductor), where pressure drives a Mott insulator to 
superconductor phase transition\cite{osc}. Thus we hypothesize
that, based on the above and overall phenomenological grounds, 
our system undergoes a 
transition from a (site diluted) Mott insulator to (site diluted) 
superconductor. The pressure in the present case is an effective 
chemical pressure, arising from reorganization of electron wave 
functions and hopping parameters, arising from a strong electronic 
coupling to the cerium subsystem.

What do we know about pressure driven Mott insulator to superconductor
transition ? The present author recently predicted \cite{gbOSC} 
such a possibility in cuprates and other systems; Zhang\cite{fczhang} has 
an independent and some what different approach.  In our mechanism long 
range coulomb interaction, originally invoked by Mott plays a key role 
in establishing a strong first order 
transition. Our key new input is to view the metallic state close to
the Mott transition as a `projected metal', a `self doped' Mott insulator, 
where superexchange survives in the conducting state. Within a simple tight 
binding model,
the transition is viewed as a spontaneous creation of a small but
finite and {\em equal density } of -e (doubly occupied orbital or
`doublon') and +e 
(empty orbital or `holon') charge carriers,
whose individual number is approximately conserved at low energies. 
What is important is that the superexchange survives in this conducting 
state - this is a memory of the Mott insulator. The concentration of 
self doped carriers 2y (= y + y  for $e^-$ and $e^+$ carriers) is decided
by long range coulomb interaction, local dielectric constant and band
parameters. A projected metal is meaningful as long as 2y $<<$ 0.5
(fig.3b). Here 0.5 is density of doubly occupied and empty orbitals in 
one orbital tight binding model of free electrons at half filling.

In our work\cite{gbOSC}, proposing a mechanism for Mott insulator to 
superconductor
transition under pressure, we introduced a 2-species t-J model, to take
care of the projected character of the self-doped state, and discussed 
superconductivity. While the insulating phase of \ndc is described by a 
Heisenberg model
on a diluted lattice, the conducting state is described by our 2-species
t-J model on a diluted lattice:

\bearr
  H_{\rm 2tJ}  & \approx &  - \sum_{ij,\sigma}t^{}_{ij}
s^{\dagger}_{i\sigma} s^{}_{j\sigma} 
( e^{}_{i} e^{\dagger}_{j}  
- d^{}_{i} d^{\dagger}_{j}) 
 + h.c. \nonumber \\
  & -  & \sum_{ij} J_{ij}
( {\bf S}_i \cdot 
{\bf S}_j - \frac{1}{4} {\rm n^{}_i n^{}_j} ) 
\eearr
including local constraint $ d^{\dagger}_{i} d^{}_{i} + e^{\dagger}_{i} e^{}_{i} +
\sum_{\sigma} s^{\dagger}_{i\sigma} s^{}_{i\sigma} = 1 $. Summation over sites
excludes `diluted' sites. Here we have
used the slave boson representation for an electron $c^{}_{i\sigma}
= e^\dagger_i s^{}_{i\sigma} + \sigma d^{}_{i} s^\dagger_{i{\bar \sigma}}$,
with e's, d's and s's representing zero occupancy (+e), double 
occupancy (-e) and single occupancy (neutral). The first term of equation 1, 
represents holon hopping and
doublon hopping, the second term the superexchange among neutral sites.
The conserved number of singly occupied sites is fixed through a global
constraint, $\sum_{i\sigma} s^{\dagger}_{i\sigma} s^{}_{i\sigma} = (1-2y)N$; 
this in turn also fixes the holon and doublon numbers to be $yN$. As mentioned
earlier, $y$ is a self consistent parameter that depends on the chemical
pressure, in our case on Ce concentration $x$. The physics of 
long range coulomb interaction that caused a first order Mott transition
is contained in our 2-species t-J model, parametrically through the
self doping density $y$.

How superconductivity arises in 2-species t-J model ? We have shown
earlier\cite{gbOSC} that for the case of bipartite lattice a 2-species 
t-J model can be exactly mapped onto the regular (single species) t-J model.
This mapping follows once we recognize that d's and e's are hard core
bosons and they also commute at different sites. The transformation 
$d_i \rightarrow \epsilon_i e_i$ implements the exact mapping. 
Here $\epsilon_i \pm$ on the A and B sublattices of the square lattice.
As site dilution preserves the bipartite character of the square
lattice, the mapping survives and we get a regular t-J model on the same 
diluted lattice.

Presence of (a small) next nearest neighbor hopping $t$' makes the 
transformation inexact. However, in such a case one can study the 
2 species t-J model directly using various known approximate methods.

What is the effect of site dilution on the symmetry of superconducting
order parameter. Site dilution is a strong perturbation. One would 
expect, based on Anderson's theorem on dirty superconductors, that 
\tc of d-wave superconductivity will be very sensitive
to the site dilution disorder. Here, something remarkable happens and 
makes site dilution disorder less effective in suppressing d-wave 
superconductivity. We will not go into the details, except to point out
that this happens within our RVB mean field analysis for the following 
reason. A general scatterer will cause phase randomization as cooper
pairs scatter in k-space, because of the change in sign (arising from
$d_{x^2-y^2}$ symmetry). However, our scatterer is a site dilution and
phase randomization does not occur, as two sublattice character
is preserved by site dilution. That is, in real space, the d-wave order 
parameter fits into a site diluted 2D square lattice without any 
frustration. 

The above is consistent with what is experimentally known about 
symmetry of the superconducting order parameter\cite{gap} and also 
the scale of superconducting \tc for the \ndc family. It should be 
pointed out that a $d_{x^2-y^2}$ to extended-S transition 
suspected\cite{sToD} in some experiments is allowed by our theory, as 
site dilution will have different quantitative effects on an extended-S 
solution and $d_{x^2-y^2}$ solution and might stablise s-wave at higher 
dopings.
\begin{figure}
\includegraphics[width=5cm]{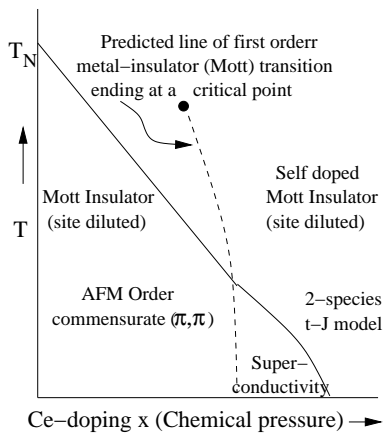}
\caption{\label{Fig2}Our proposed phase diagram. A key prediction 
is the presence of a line of first order metal-insulator (Mott) transition 
ending at a critical point}
\end{figure}

In the light of our model the following experimental results get
organized and are less puzzling, and also some predictions follow.
We summarize them below:

i) In a recent work, Tsukada et al.,\cite{Naito} found superconductivity 
in \lco in T'-structure, for trivalent rare earth substitution, \lar; 
R =  Lu and Y, among others. This is an unexpected and remarkable 
finding. A well known stable trivalent state of Lu and Y makes 
doping of CuO$_2$  plane unlikely. We suggest that \lco in T'-structure
has an inherent enhanced chemical pressure, which is further increased by
Lu or Y substitution leading to Mott insulator superconductor
transition without external doping, via our mechanism.
\begin{figure}
\includegraphics[width=8cm]{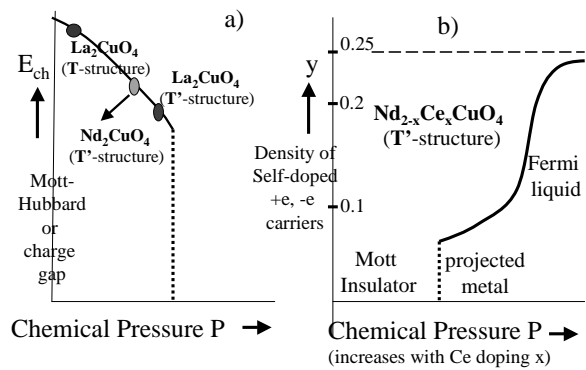}
\caption{\label{Fig3} 
a) Mott-Hubbard or Charge gap $E_c$ as a function of chemical pressure.
Schematically, three experimentally studied systems have different
charge gaps as shown, b) Self-doped carrier density $y$ as a function 
of chemical pressure (related to Ce doping $x$). Self doping looses its
meaning as $y$ approaches 0.25, and we leave a projected metal
and enter a fermi liquid like state}
\end{figure}

A recent inelastic X-ray scattering study\cite{MHgap} gives a Mott 
Hubbard gap of \ndc smaller than \lsco by about 1 eV. 
This is an experimental
evidence for different quantum chemistry in T and T' structure
and that \ndc is under higher chemical pressure and close to a Mott 
transition point compared to \lco (figure 3a). It will be important
to measure the Mott Hubbard gap in \lco in T'-structure, which 
according our suggestion will also be close to the Mott transition
point.

ii) The shape of the superconductor-antiferromagnetic phase boundary and 
various experiments\cite{phaseSep} indicate a strong first order character 
of the transition and phase separation. Our hypothesis B not only
explains it in a most natural fashion but also makes the following
important prediction. On very general thermodynamic grounds, 
the first order transition (as it is driven by coulomb interaction
among charges) will continue beyond the antiferromagnetic phase 
boundary into the paramagnetic normal state and end at a critical 
point, as shown in figure 2. This is experimentally established
in organic superconductors, where pressure drives a Mott insulator
to superconductor transition.

When the critical concentration $x_c$ is small, disorder will 
have less effect on the first order line and the end critical point. 
As smaller critical $x_c$ is realized in annealed \prc and \lco systems,
it may  be easier to confirm our prediction of first order transition
line and critical point in the normal state. Nuclear and muon spin resonance
will be good local probes to study the above. Further, as the first order 
phase transition is charge driven, there will be a volume change across the
phase boundary, due to coupling of the charges with the lattice. This can
be detected by sound velocity, elastic constant studies. 

\begin{figure}
\includegraphics[width=7cm]{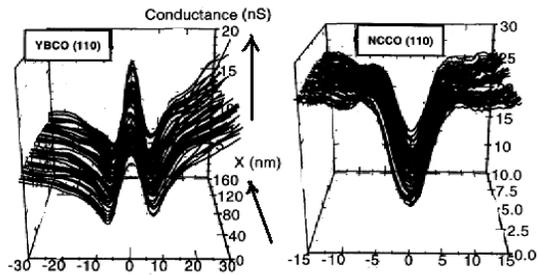}
\caption{\label{Fig1}
STM spectra, reproduced from reference\cite{tunneling1}, for 
various tip to surface distance X. According to our theory, \ndc, 
which is self-doped with equal density of +e and -e carriers, has 
a symmetric spectra.
Hole doped YBCO, which has only +e carriers has an asymmetric spectra.
x-axis scale is meV}
\end{figure}
iii) In a recent paper, Anderson and Ong\cite{pwaOng} have suggested that 
tunneling conductance should be intrinsically asymmetric, in either
electron or hole doped Mott insulators. We have applied their analysis 
to our 2-species t-J model and find a symmetric tunneling spectra, 
as expected on symmetry grounds. There is some asymmetry to the extent
of particle-hole asymmetry (arising from t' etc.). Recent results 
show\cite{tunneling1,tunneling2} a high symmetry in the tunnelling 
conductance. In figure 4, the high symmetry in \ndc compared to the 
case of YBCO of reference\cite{tunneling1} is exhibited; it is consistent 
with our proposal.  To substantiate our model further, it will be interesting 
to perform more study on symmetry aspect of tunneling in \ndc, as well as 
other systems with T' structure.

iv) Earlier electrical transport studies have invoked\cite{transport}
 the presence of two 
species of charges to explain certain anomalous results. It will be 
interesting to look for our prediction of equal density of hole and 
electron carriers in the conducting state of \ndc family and other 
T' structures, using other experiments.

v) There has been extensive ARPES\cite{ARPES} results on fermi surface 
measurements in \ndc family. ARPES finds a Luttinger volume of 
$(1-x)N\over 2$ k-points. This is consistent with our model - 
{\em our conducting half filled band, in view of site dilution, 
has same number of effective k-points as a full lattice with
an electron doping density of $x$}.)

vi) Recent NMR results and other experiments\cite{kitaoka}
indicate a `over doped' fermi
liquid like behavior in the normal state, We believe that this is due
to the fact that our effective carrier density 2y is beyond the standard
optimal doping density of 0.15, making it effectively over doped. 
As we discussed earlier, Ce density x is different from self doping 
density 2y; and 2y is determined by band parameters, dielectric constant 
and coulomb interaction physics.

The author wishes to thank P W Anderson, M. Fujita, T. Sato and T Tohyama 
for discussions, Dr M Naito for a correspondence and H Fukuyama and 
S Maekawa for hospitality at IFCAM, IMR, Tohoku University, where part 
of this work was carried out.

\end{document}